\documentclass{emulateapj}

\usepackage{natbib}
\usepackage{amsmath}

\newcommand{\about}{$\sim\!\!$~}
\newcommand{\kms}{\,km\,s$^{-1}$}

\newcommand{\msun}{M$_{\sun}$}
\def\lsim{\hbox{\rlap{\raise 0.425ex\hbox{$<$}}\lower 0.65ex\hbox{$\sim$}}}
\def\gsim{\hbox{\rlap{\raise 0.425ex\hbox{$>$}}\lower 0.65ex\hbox{$\sim$}}}

\def\arcdeg{\mbox{$^\circ$}}

\shorttitle{Early/Late Observations of SN~2008ha}
\shortauthors{Foley et~al.}

\begin{document}

 \title{Early and Late-Time Observations of SN~2008ha: Additional
 Constraints for the Progenitor and Explosion}

\def\cfa{1}
\def\clay{2}
\def\psu{3}
\def\harv{4}
\def\pitt{5}

\author{
{Ryan~J.~Foley}\altaffilmark{\cfa,\clay},
{Peter~J.~Brown}\altaffilmark{\psu},
{Armin~Rest}\altaffilmark{\harv},
{Peter~J.~Challis}\altaffilmark{\cfa},
{Robert~P.~Kirshner}\altaffilmark{\cfa}, and
{W.~Michael~Wood-Vasey}\altaffilmark{\pitt}
}

\altaffiltext{\cfa}{
Harvard-Smithsonian Center for Astrophysics,
60 Garden Street, 
Cambridge, MA 02138.
}
\altaffiltext{\clay}{
Clay Fellow. Electronic address rfoley@cfa.harvard.edu .
}
\altaffiltext{\psu}{
Pennsylvania State University,
Department of Astronomy \& Astrophysics,
University Park, PA 16802.
}
\altaffiltext{\harv}{
Department of Physics,
Harvard University,
17 Oxford Street,
Cambridge, MA 02138
}
\altaffiltext{\pitt}{
Department of Physics and Astronomy,
100 Allen Hall,
3941 O'Hara St,
University of Pittsburgh,
Pittsburgh, PA 15260.
}

\begin{abstract}
We present a new maximum-light optical spectrum of the the extremely
low luminosity and exceptionally low energy Type Ia supernova (SN~Ia)
2008ha, obtained one week before the earliest published spectrum.
Previous observations of SN~2008ha were unable to distinguish between
a massive star and white dwarf origin for the SN.  The new
maximum-light spectrum, obtained one week before the earliest
previously published spectrum, unambiguously shows features
corresponding to intermediate mass elements, including silicon,
sulfur, and carbon.  Although strong silicon features are seen in some
core-collapse SNe, sulfur features, which are a signature of
carbon/oxygen burning, have always been observed to be weak in such
events.  It is therefore likely that SN~2008ha was the result of a
thermonuclear explosion of a carbon-oxygen white dwarf.  Carbon
features at maximum light show that unburned material is present to
significant depths in the SN ejecta, strengthening the case that
SN~2008ha was a failed deflagration.  We also present late-time
imaging and spectroscopy that are consistent with this scenario.
\end{abstract}


\defcitealias{Valenti09}{V09}
\defcitealias{Foley09:08ha}{Paper~I}


\section{Introduction}\label{s:intro}

Supernova (SN) 2008ha has the lowest peak luminosity and ejecta
velocity of any SN~Ia yet observed \citep[hereafter Paper I and V09,
respectively]{Foley09:08ha, Valenti09}.  It peaked at $M_{V} =
-14.2$~mag, had an ejecta velocity a week after maximum of \about
2000~\kms, and had a rise time of \about 10~days; these values are
\about 5 magnitudes fainter, 8000~\kms\ slower, and 10~days shorter
than that of normal SNe~Ia, respectively.  Together, these
observations indicate a very low kinetic energy of \about $2 \times
10^{48}$~ergs and an ejecta mass of 0.15~\msun\
\citepalias{Foley09:08ha}.

Although SN~2008ha resembles SN~2002cx, the prototype of a class of
peculiar SNe~Ia with lower than normal luminosity and ejecta velocity
(see \citealt{Jha06:02cx} for a review of the class), the similar
energetics and spectra of SN~2008ha and some peculiar core-collapse
SNe led \citetalias{Valenti09} to suggest that the progenitor of
SN~2008ha was a massive star.  In \citetalias{Foley09:08ha} we
considered several progenitor and explosion models: (1) the collapse
of a massive star to a black hole with most of the star ``falling
back''; (2) a runaway nuclear reaction caused by electron capture on
a white dwarf (WD); (3) the failed deflagration of a WD; and (4)
nuclear burning of a massive He shell on the surface of a WD in an AM
CVn system (a ``SN~.Ia''; \citealt{Bildsten07}).

Delicate balancing of the energetics and the detection of a
SN~2002cx-like object in an S0 galaxy makes the fallback model
unfavorable for SN~2008ha and the class, respectively.  The
electron-capture scenario predicts complete burning to $^{56}$Ni with
no intermediate-mass elements (IMEs).  The observations of SN~2008ha
and other members of the SN~2002cx-like class are all consistent with
a deflagration (with SN~2008ha being a failed deflagration;
\citetalias{Foley09:08ha}).  The .Ia model fits SN~2008ha, but can not
currently reproduce the luminosity required for other members of the
class.

Nucleosynthetic models provides additional information to distinguish
between these models.  Explosive C/O burning produces a significant
amount of S while He burning produces more Ca (see
\citealt{Perets09} for a recent examination of explosions with
different fractions of each process).  Furthermore, it is expected
that S is contained in the inner layers of a core-collapse explosion,
while it is predominantly in the outer layers of a thermonuclear
explosion \citep[e.g.,][]{Thielemann91}.  \citetalias{Valenti09}
argued that the lack of S in their first spectrum was most consistent
with a massive star origin.  A strong detection of S at early times
would give support for C/O burning and a WD progenitor.

The SN~2008ha spectra presented in \citetalias{Foley09:08ha} and
\citetalias{Valenti09} spanned the phases of 6.5 to 68.1~days past
maximum light in the $B$ band (2008 November 12.7 UT;
\citetalias{Foley09:08ha}; UT dates will be used throughout this
paper).  Here we present two spectra which extend this range in both
directions, with phases of $-1$ and 231~days relative to $B$ maximum.
We also present late-time photometry which constrains the amount of
generated $^{56}$Ni.  These observations, including the detection of
strong S lines in the maximum-light spectrum, further constrain the
possible progenitor and explosion models for SN~2008ha, and highly
favor a WD progenitor.


\section{Observations and Data Reduction}\label{s:obs}

We have obtained two new spectra of SN~2008ha.  The early-time
spectrum was obtained on 2008 November 11 (1~day before $B$ maximum)
with the Hobby-Eberly Telescope using the Low Resolution Spectrograph
\citep{Hill98}.  The spectrum is the combination of four
450-second exposures.  The late-time spectrum was obtained on 2009
July 1/2 (231~days after $B$ maximum) with Gemini-North using GMOS in
nod and shuffle (N\&S) mode \citep{Hook04}.  The GMOS spectrum is the
combination of six 1280-second (after accounting for nod time)
exposures, producing 12 spectra with effective exposure times of
640~seconds and a total exposure time of 7680~seconds.  The GMOS
spectra were obtained with three slightly different central
wavelengths to compensate for chip gaps.  Five of the six exposures
were obtained on July 1, with the remaining exposure obtained on July
2.

Standard CCD processing and spectrum extraction were accomplished with
IRAF\footnote{IRAF: the Image Reduction and Analysis Facility is
distributed by the National Optical Astronomy Observatory, which is
operated by the Association of Universities for Research in Astronomy,
Inc.\ under cooperative agreement with the National Science
Foundation.}.  The GMOS data were reduced using the Gemini IRAF N\&S
package.  Low-order polynomial fits to calibration-lamp spectra were
used to establish the wavelength scale, and small adjustments derived
from night-sky lines in the object frames were applied.  We employed
our own IDL routines to flux calibrate the data and remove telluric
lines using the well-exposed continua of spectrophotometric standards
\citep{Wade88, Foley03, Foley06}.

Two epochs of $r$-band images were obtained on 2009 June 23 and 2009
July 1/2 with
LDSS3\footnote{http://www.lco.cl/telescopes-information/magellan/instruments-1/ldss-3-1/
.} on the Magellan Clay telescope and with GMOS on the Gemini-North
telescope, respectively.  These images were reduced using standard
techniques.  The flux of the SN was determined by fitting a PSF
profile in the flattened images using the DoPHOT photometry package
\citep{Schechter93}.  Since all of our images presumably have some SN
flux, difference imaging is not yet possible.  The presence of an
underlying \ion{H}{2} region causes our flux measurements to be upper
limits to the true SN flux.  We convert the $r$-band instrumental
magnitudes into $R$-band magnitudes using the $R$-band catalog magnitudes of
local standard stars \citepalias{Foley09:08ha}.  No instrumental color
corrections were applied.


\section{Spectral Analysis}\label{s:spec}

\subsection{Maximum-Light Spectrum}

\begin{figure}
\begin{center}
\epsscale{1.25}
\rotatebox{90}{
\plotone{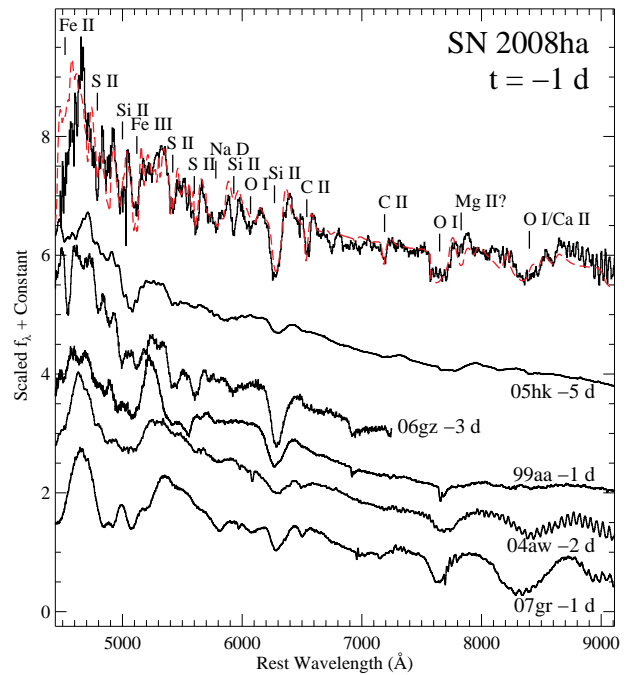}}
\caption{Optical spectrum of SN~2008ha at $t = -1$~days relative to
$B$ maximum (solid black line) and a SYNOW model fit (red dashed
line).  The spectrum is dominated by IMEs.  Spectra of SNe~1999aa,
2004aw, 2005hk, 2006gz, and 2007gr are shown for comparison (each
shifted by selected amounts to give the best match to the ejecta
velocity of SN~2008ha).}\label{f:spec_early}
\end{center}
\end{figure}

The maximum-light spectrum (shown in Figure~\ref{f:spec_early})
consists almost exclusively of IMEs with low ejecta velocities.  The
minima of the \ion{Si}{2} $\lambda 6355$ and \ion{O}{1} $\lambda 7774$
features are blueshifted by \about 3700 and 5100~\kms, respectively.
The \ion{Si}{2} feature was much weaker a week later (the first
spectrum from \citetalias{Foley09:08ha}) and could not be directly
measured, but the characteristic photospheric velocity of \about
2000~\kms\ for the 6~day spectrum is approximately half that of
\ion{Si}{2} in the maximum-light spectrum.  Similarly the \ion{O}{1}
velocity is 2.5 times larger at maximum brightness than it is at $t =
6$~days.

The identification of IMEs is confirmed by fitting the spectrum with
the SN spectrum-synthesis code SYNOW \citep{Fisher97}.  Although SYNOW
has a simple, parametric approach to creating synthetic spectra, it is
a powerful tool to aid line identifications (see
\citetalias{Foley09:08ha} for details).  In the early-time spectrum,
SYNOW unambiguously identifies the presence of O, Na, Si, S, Ca, and
Fe, all common features in maximum-light SN~Ia spectra, as well C,
which is rarely seen.  Although strong \ion{Si}{2} is the hallmark
feature of SNe~Ia, there are examples of SNe~Ic with strong Si
features \citep{Taubenberger06, Valenti08}; however, those objects did
not have strong \ion{S}{2} features.

The SYNOW fit of SN~2008ha is similar to those of the $-1$ and
$-5$~day spectra of SNe~2002cx \citep{Branch04} and 2005hk \citep[a
SN~2002cx-like object;][]{Chornock06}, respectively.  The differences
are primarily a lower expansion velocity and the presence of
\ion{C}{2} for SN~2008ha (\citealt{Chornock06} did fit the SN~2005hk
spectrum with \ion{C}{3}, but the identification was ambiguous).
SN~2008ha appears to have weaker Fe lines than the other objects,
signaling a lower $^{56}$Ni yield relative to that of IMEs.

We compare the maximum-light spectra of SN~2008ha to SNe~Ia and SNe~Ic
finding that SN~2008ha most closely resembles a SN~Ia.  Our comparison
spectra (see Figure~\ref{f:spec_early}) include SN~2005hk, a SN~Ia
similar to SN~2002cx \citep{Phillips07}, SN~1999aa, a slightly
over-luminous SN~Ia \citep{Garavini04}, SN~2006gz, a luminous SN~Ia
with C features in its pre-maximum spectra \citep{Hicken07},
SN~2004aw, a SN~Ic that was originally classified as a SN~Ia (its true
nature was only realized with nebular spectra;
\citealt{Taubenberger06}), and SN~2007gr, a well-observed SN~Ic with a
spectrum somewhat similar to SN~2004aw \citep{Valenti08}.  The
maximum-light spectrum of SN~2008ha is generally similar to all five
comparison spectra after correcting for velocity differences.  But the
SNe~Ic have weaker \ion{Si}{2} than that seen in SN~2008ha and lack
obvious \ion{S}{2} lines.  The SNe~Ia (and particularly SN~2006gz)
have spectra that are more similar to the spectrum of SN~2008ha,
strengthening the classification of SN~Ia for SN~2008ha; however, the
spectra still have differences, and it is difficult to definitively
place SN~2008ha in this class.

The spectrum of SN~2008ha unambiguously contains the signature of C in
its ejecta, showing strong absorption corresponding to \ion{C}{2}
$\lambda 6580$ and \ion{C}{2} $\lambda 7234$.  Carbon lines have been
seen in the spectra of some SNe~Ia \citep[e.g.,][]{Hicken07} and in
some SNe~Ic \citep[e.g.,][]{Valenti08}.  Although it is rare for
SNe~Ia to show strong C absorption near maximum light, there is an
example in the small sample \citep{Yamanaka09}.  It is worth noting
that there are also SNe~Ic with C features at this phase.  If
SN~2008ha was the result of a WD explosion, a significant amount of C
extends further into the ejecta than most SNe~Ia.  If the progenitor
of SN~2008ha was a massive star, then the strong C lines at maximum
light may indicate a large C layer or significant mixing in the
progenitor.

\subsection{Late-Time Spectrum}

\begin{figure}
\begin{center}
\epsscale{1.55}
\rotatebox{90}{
\plotone{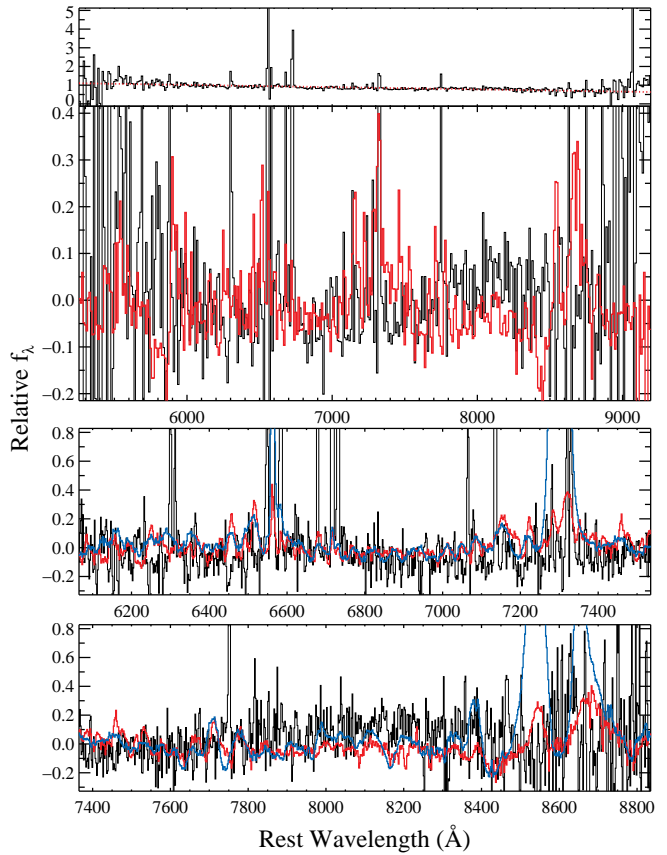}}
\caption{({\it top panel}): The extracted
231~day optical spectrum of SN~2008ha (solid black line), which is
contaminated by an underlying \ion{H}{2} region.  A linear fit to the
continuum is also shown (red dotted line).  ({\it second panel}): The
continuum-subtracted spectra of SNe~2008ha (black line) and 2002cx ($t
= 227$~days; red line; \citealt{Jha06:02cx}).  ({\it bottom panels}):
The continuum-subtracted spectrum of SN~2008ha from above is
reproduced on an enlarged scale with finer resolution.  The red and
blue lines are the continuum-subtracted 227~day spectrum of SN~2002cx
and 62~day spectrum of SN~2008ha \citepalias{Foley09:08ha}.
}\label{f:spec_late}
\end{center}
\end{figure}

The late-time ($t = 231$~days) spectrum of SN~2008ha is shown in
Figure~\ref{f:spec_late} along with the 227-day spectrum of SN~2002cx.
We compensate for the contamination of an underlying \ion{H}{2} region
by subtracting a linear continuum fit from each SN spectrum and
compare the residual spectra.

The lower panels of Figure~\ref{f:spec_late} present the
continuum-subtracted spectrum of SN~2008ha on an expanded wavelength
scale.  The low S/N of the spectrum prevents us from
detecting any individual feature in a SN~2002cx-like spectrum other
than possibly [\ion{Ca}{2}] $\lambda \lambda 7291$, 7324 or the
\ion{Ca}{2} NIR triplet (8498, 8542, and 8662~\AA).
The late-time spectrum of SN~2008ha exhibits no obvious emission
from [\ion{O}{1}] (which is expected for SNe~Ic and predicted for
deflagration models; \citealt{Gamezo03}), [\ion{Ca}{2}], or
\ion{Ca}{2}.
In fact, there appears to be absorption at the position that we would
expect \ion{Ca}{2} emission.  As the observations were performed in
N\&S mode, there was no local background subtraction performed and
these features are not an artifact of having SN light in a background
region.  This wavelength range has significant night sky emission, and
imperfect sky-line subtraction may cause such features.
Regardless, there are no
definitive detections of strong, broad features corresponding to
\ion{Ca}{2}.

Between $t = 56$ and 227~days, the velocity width of the [\ion{Ca}{2}]
feature in SN~2002cx decreased from FWHM$\ \approx 2400$~\kms\ to
\about 900~\kms.
In the 62~day spectrum of SN~2008ha, [\ion{Ca}{2}] had FWHM$\ \approx
900$~\kms.  If the width of the feature decreases at the same rate in
both object, we expect the feature to have FWHM$\ \approx 350$~\kms\
in the late-time spectrum of SN~2008ha.  All narrow lines have a
similar redshift and velocity width (\about 200~\kms\ FWHM, equivalent
to the instrumental resolution).  It is possible that the velocity
width is below our instrumental resolution, and in that case we could
mistake lines from the SN as lines from the \ion{H}{2} region (such as
[\ion{O}{1}]).  There are no intermediate-width emission lines (such
as H or He) that one would expect if the SN ejecta were interacting
with a dense circumstellar medium associated with a massive star
progenitor.

The ratio of [\ion{N}{2}] $\lambda 6584$ and H$\alpha$ lines from the
\ion{H}{2} region is 0.046, corresponding to a metallicity (as
determined by the N2 method of \citealt{Pettini04}) of $12 +
\log({\rm 0}/{\rm H}) = 8.14 \pm 0.03$ (stat).  This is consistent
with that of a nearby \ion{H}{2} region \citepalias{Foley09:08ha} and
indicative of no significant H emission from circumstellar
interaction.


\section{Photometric Analysis}\label{s:phot}

\begin{figure}
\begin{center}
\epsscale{1.2}
\rotatebox{90}{
\plotone{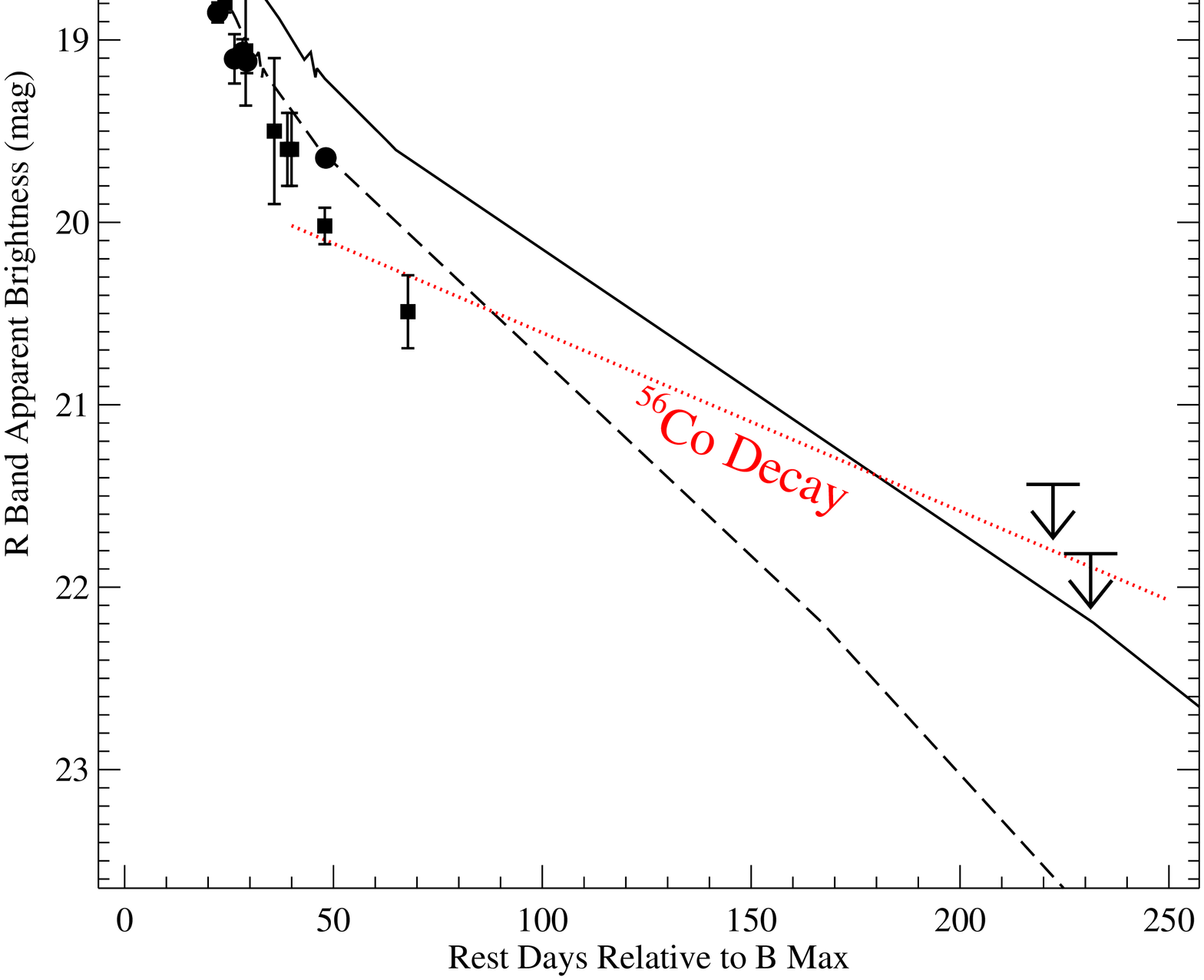}}
\caption{$R$-band light curve of SN~2008ha.  The circles and squares
are from \citetalias{Foley09:08ha} and \citetalias{Valenti09},
respectively.  The upper limits are from our new photometry.  The red
dotted line indicates the decay expected from generating $10^{-3}
M_{\sun}$ of $^{56}$Ni.  The solid and dashed black lines is the light
curve of SN~2005hk \citep{Phillips07, Sahu08} shifted to match the
peak brightness of SN~2008ha.  The dashed line has been scaled to
match the width of SN~2008ha (see \citetalias{Foley09:08ha} for
details), while the solid line has not been modified.}\label{f:lc}
\end{center}
\end{figure}

We reproduce the early-time light curves of SN~2008ha from 
\citetalias{Foley09:08ha} and \citetalias{Valenti09} in
Figure~\ref{f:lc} along with our two late-time photometric upper
limits.  Our late-time imaging resulted in $R$-band measurements of
$21.43 \pm 0.05$ and $21.82 \pm 0.05$~mag 222 and 231~days after $B$
maximum, respectively.  However, these images have had no template
subtraction, so the photometry reflects the brightness of the
combination of the SN and the underlying \ion{H}{2} region, and
therefore, these numbers are upper limits on the brightness of the SN.
The differences in two measurements may indicate real fading between
the two epochs, but can also be explained as systematic differences in
the instrument/filter responses.

\citetalias{Foley09:08ha} found that the light curves of SNe~2005hk
and 2008ha were well matched if the light curve of SN~2005hk was
``stretched'' by a factor of 0.73.  In Figure~\ref{f:lc}, we compare
the $R$-band light curve of SN~2008ha to that of SN~2005hk
\citep{Phillips07, Sahu08} both with and without this stretching.  The
stretched light curve is a good approximation of the early-time
behavior because the SEDs are similar.  But the light curves exhibit
different decline rates because the SNe had different opacities,
amounts of ejecta/$^{56}$Ni, and kinetic energies.  The ejecta of
SN~2002cx-like objects appear to be optically thick at very late times
\citep{Jha06:02cx, Sahu08}, so opacity effects may still dominate at
$t = 250$~days.  Regardless, the SN~2005hk light curves should give an
indication of the expected decay rate at late times if SN~2008ha
evolves in a similar fashion.

If the peak luminosity is powered by $^{56}$Ni decay, then the
luminosity at late times should be powered by $^{56}$Co (the decay
product of $^{56}$Ni) decay.  The brightness at late times is
therefore directly related to the $^{56}$Ni mass.  Assuming the
relationship between $^{56}$Ni mass and bolometric luminosity
\citep{Sutherland84}, a solar SED for bolometric corrections, and full
$\gamma$-ray trapping, the late-time photometry places a limit of
$M_{^{56} {\rm Ni}} \lesssim 10^{-3} M_{\sun}$ (see
Figure~\ref{f:lc}).  This value is also consistent with the brightness
of SN~2008ha at $t \approx 60$~days.  Considering the uncertainty of
the SED, this is consistent with the estimate from the early-time
light curve of $(3.0 \pm 0.9) \times 10^{-3} M_{\sun}$.


\section{Discussion and Conclusions}\label{s:disc}

The maximum-light spectrum provides key information for understanding
the nature of SN~2008ha.  The spectrum has strong lines from IMEs,
including Si and S.  Nucleosynthetic models suggest that strong S
features are indicative of C/O burning \citep[e.g.,][]{Perets09}.
Furthermore, no core-collapse SN has been observed to have strong S
lines at maximum light, and having these features in the maximum-light
spectrum indicates that a non-negligible amount of S is in the outer
ejecta, suggesting a WD progenitor \citep[e.g.,][]{Thielemann91}.  One
of the points adduced by \citetalias{Valenti09} in favor of the
core-collapse interpretation of this event was the absence of
\ion{S}{2} and the weak \ion{Si}{2} seen in their earliest spectrum,
8~days after maximum brightness.  The new data presented here show
that \ion{S}{2} is present and \ion{Si}{2} is significantly stronger
than in those data.  Although peculiar abundances in the outer layers
of a massive star may be the cause of the S lines, the data currently
favor the idea of a WD progenitor.

SN~2008ha shares many similarities (luminosity, spectral features,
etc) with SN~2005E, a low-luminosity SN~Ib with strong Ca lines at
late times (\citetalias{Foley09:08ha}; \citealt{Perets09}.  There have
been several SN~2005E-like objects discovered, but they exist
predominantly in early-type galaxies, in contrast to the primarily
late-type hosts of SN~2002cx-like objects (SN~2008ha was discovered in
an dwarf irregular galaxy), and suggestive that they have old
progenitors.  \citet{Perets09} found that by changing the amount of He
and C/O burning, the ratio of S-to-Ca in the ejecta of a SN can be
manipulated.  SNe~2005E and 2008ha may have similar progenitors and/or
explosion mechanisms.  The different host-galaxy populations of these
classes may indicate that progenitor age or metallicity has an affect
on the resulting explosion, similar to SN~1991T and SN~1991bg-like
objects.

To derive the previous estimates of the ejecta mass and kinetic
energy, the velocity at 6 days past maximum brightness was
extrapolated to the time of maximum assuming a velocity gradient
similar to that of a normal SN~Ia.  With these new data, the
extrapolation is not necessary, and the systematic uncertainty
related to these measurements can be reduced.  The velocity of the
maximum-light spectrum is twice that of the adopted value from
\citetalias{Foley09:08ha}, increasing the kinetic energy and ejecta
mass by factors of 8 ($E \propto v^{3}$) and 2 ($M \propto v$),
respectively.  We therefore revise our estimates of the kinetic energy
to be $1.8 \times 10^{49}$~ergs and $M_{\rm ej} = 0.3 M_{\sun}$.  We
note that this analysis assumes that the composition and opacity of
the ejecta of SN~2008ha are similar to those of a normal SN~Ia, which
may cause systematic uncertainty of order a factor of two (see
\citetalias{Foley09:08ha} for details).

The late-time photometry is consistent with the production of a few
times $10^{-3} M_{\sun}$ of $^{56}$Ni, similar to estimates from the
early-time light curve \citepalias{Foley09:08ha, Valenti09}.  The
late-time spectrum shows that there are no extremely strong emission
lines from the SN; however, the relatively low S/N spectrum places
weak limits on such features.

The strong C lines in the maximum-light spectrum indicates that there
is unburned material far into the ejecta, consistent with a
deflagration \citep{Gamezo03}.  The low ejecta and $^{56}$Ni mass are
consistent with a failed deflagration of a WD that did not disrupt the
progenitor \citepalias{Foley09:08ha}, but are perhaps also explained
by a sub-Chandrasekhar mass or more exotic WD explosion.

Pre-maximum data is critical for understanding the nature of
SN~2008ha-like objects.  Although our observations are all consistent
with a failed deflagration of a WD, we can not completely rule out
other models.  If the progenitor of SN~2008ha was a very massive star,
one might expect a brightening in X-rays or radio if the progenitor
had a wind or pre-explosion outbursts.  Future early-time X-ray and
radio observations will help constrain the nature of similar events.
Eventually we will detect a SN~2008ha-like object with deep
pre-imaging data and either detect or highly constrain the properties
of the progenitor.

\begin{acknowledgments} 

R.J.F.\ is supported by a Clay Fellowship.  R.J.F.\ would like to
thank G.\ Narayan and B.\ Stalder for managing the imaging data, and
the Magellan and Gemini staffs for being extremely accommodating.
Discussions during the KITP conference ``Stellar Death and
Supernovae'' were beneficial to this study.  Supernova research at
Harvard is supported by NSF grant AST09--07903.

Based in part on observations obtained at the Gemini Observatory,
which is operated by the Association of Universities for Research in
Astronomy, Inc., under a cooperative agreement with the US National
Science Foundation on behalf of the Gemini partnership: the NSF
(United States), the Science and Technology Facilities Council (United
Kingdom), the National Research Council (Canada), CONICYT (Chile), the
Australian Research Council (Australia), Minist\'{e}rio da Ci\^{e}ncia
e Tecnologia (Brazil) and Ministerio de Ciencia, Tecnolog\'{i}a e
Innovaci\'{o}n Productiva (Argentina).  The HET is a joint project of
the University of Texas at Austin, the Pennsylvania State University,
Stanford University, Ludwig-Maximilians-Universit\"{a}t M\"{u}nchen,
and Georg-August-Universit\"{a}t G\"{o}ttingen. The HET is named in
honor of its principal benefactors, William P. Hobby and Robert
E. Eberly.  The Marcario LRS is named for Mike Marcario of High
Lonesome Optics who fabricated several optics for the instrument but
died before its completion.  The LRS is a joint project of the HET
partnership and the Instituto de Astronom\'{i}a de la Universidad
Nacional Aut\'{o}noma de M\'{e}xico.

{\it Facilities:} 
\facility{Gemini:North(GMOS), HET(LRS), Magellan:Clay(LDSS3)}

\end{acknowledgments}



\end{document}